# An Optimal Game Approach for Heterogeneous Vehicular Network Selection with Varying Network Performance

Xiangmo Zhao, Xiaochi Li, Zhigang Xu, and Ting Chen

*Abstract*—**Most conventional heterogeneous network selection strategies applied in heterogeneous vehicular network regard the performance of each network constant in various traffic scenarios. This assumption leads such strategies to be ineffective in the real-world performance-changing scenarios. To solve this problem, we propose an optimal game approach for heterogeneous vehicular network selection under conditions in which the performance parameters of some networks are changing. Terminals attempting to switch to the network with higher evaluation is formulated as a multi-play non-cooperative game. Heterogeneous vehicular network characteristics are thoroughly accounted for to adjust the game strategy and adapt to the vehicular environment for stability and rapid convergence. A multi-play non-cooperative game model is built to formulate network selection. A probabilistic strategy is used to gradually drive players toward convergence to prevent instability. Furthermore, a system prototype was built at the Connected and Automated Vehicle Test bed of Chang'an University (CAVTest). Its corresponding test results indicate that the proposed approach can effectively suppress the ping-pong effect caused by massive handoffs due to varying network performance and thus well outperforms the single-play strategy.**

*Index Terms*—**Connected Vehicle, Dedicated Short-Range Communication, Heterogeneous Network, Intelligent Transportation System**

## I. Introduction

Heterogeneous network has gathered a great deal of attention in the field of Intelligent Transportation Systems (ITS) [1], because there is no one network that can meet the myriad demands of vehicular applications. TABLE I shows a comparison between vehicular network demands and the performance of a single network [2]–[4]. As the most mature and widely used network in vehicular environments, Dedicated Short-Range Communication (DSRC) transmits information by broadcasting. This effectively minimizes propagation delay and adapts to the rapidly changing network topology [2]. However, DSRC will encounter harsher conditions in the real-world scenario than it was designed for. Assume that the transmit range of DSRC is 300 m and the maximum capability of DSRC is 50, DSRC cannot load all broadcasting information in a full-volume four-lane or six-lane road with a traffic system that achieves maximum at density of 30 veh/km. Severe packet loss occurs when there are more than 50 terminals because of DSRC's ad hoc framework and Media Access Control (MAC) protocol [5]. The limited bandwidth of DSRC also prevents it from loading the latest large-capacity multimedia applications [6]. On the other hand, although Long Term Evolution (LTE) can meet most requirements of various vehicular network applications from the design point of view [7], its performance often significantly degrades because of network optimization, blocking, interference, and other abruptly changed conditions in actual deployment [6]. Commercial Wi-Fi can provide high-throughput network service with low latency; however, the transmitting range is not wide enough and horizontal handoff consumption is too much for vehicular network. These issues bring critical challenges to many safety-related ITS applications, such as Vehicle-to-Vehicle (V2V) based Forward Collision Warning (FCW), Cooperative Adaptive Cruise Control (CACC), and so on. So, there is an urgent demand for integration of different access networks to make them fully complementary [8].

The goal of regular heterogeneous network is to ensure that terminals are best connected, anytime and anywhere [9]. Compared to regular heterogeneous telecom network, the fusion of heterogeneous vehicular networks needs to meet many stringent demands for network selection because of the following reasons: (1) crowded moving on-board terminals, (2) difficulty in centralized decisions, and (3) degraded network performance might incur critical issues related to life and property loss. The vehicular network system must be designed for utmost reliability. The network performance is highly sensitive to data propagation delay and Packet Loss Rate (PLR). These facts make traditional vertical handoff strategies ineffective in heterogeneous vehicular network. During the past 20 years, many researchers have investigated vertical handoff

This work was supported by the National Natural Science Foundation of China (No. 51278058), Joint Laboratory of Internet of Vehicles sponsored by Ministry of Education and China Mobile (No. 213024170015), and 111 Project(B1403).

X. Zhao, X. Li, Z. Xu, and T. Chen are with the School of Information Engineering, Chang'an University, Xi'an 710064, China (e-mail: xmzhao@chd.edu.cn; xcli@chd.edu.cn; xuzhigang@chd.edu.cn; tchen@chd.edu.cn). Corresponding author: xuzhigang@chd.edu.cn , xcli@chd.edu.cn;

TABLE I
COMPARISON BETWEEN VEHICULAR DEMANDS AND VEHICULAR NETWORK PERFORMANCE

| Items | Demands | DSRC | LTE [c] | Wi-Fi [d] |
|---|---|---|---|---|
| Propagation delay | Under 100ms | Around 10 ms | Around 100 ms | Depending on different environment |
| Data rate | Depending on apps | 27 MBps | Downlink:100 MBps/ Uplink:50 MBps | 600 MBps |
| Density | 240 OBUs (On-Board Units) per km [a] | 83 OBUs per km [b] | Depending on layout of base station | Depending on layout of access point |
| Mobility | 120 km/h or higher | 120 mile/h, i.e., about 196 km/h | 0-15 km/h: optimized for low mobile speed. 12-120 km/h: high performance for higher mobile speed. | No mobility support in detail |
| Horizontal handoff delay | Under 100 ms as it is demanded by propagation delay | No need to horizontal handoff | Less than 300 ms interruption time | No common ways |

[a]Density demands embody a congested two-way six-lane suburban highway with vehicles at speed of 36 km/h.
[b]DSRC density support is calculated when the broadcast range of DSRC is 300 m and maximum number of terminals is 50 in the broadcast area.
[c]LTE performance is presented as requirements for 3GPP LTE. In the real-world scenario, it is difficult to realize these requirements.
[d]Wi-Fi is not designed for vehicular networks or large-scale outdoor connections, so the radio transmission distance is relatively short. We choose IEEE 802.11n as the representative protocol per its longest range of 250 m among all Wi-Fi protocols.

strategies in heterogeneous vehicular network, such as game theory, Multi-Criteria Decision Making (MCDM), which provide always-on connectivity and optimized performance with regard to user preference, application requirements, and other contexts [10].

Most vertical handoff strategies applied to heterogeneous vehicular network assume network performance as a constant parameter. In real application scenario, the network performance constantly changes due to network optimization, barriers, noise, and especially, the changing number of terminals. Vinel [5] found that the number of terminals dominates the network performance for both LTE and DSRC when it becomes excessive within the selected area. This may cause massive handoffs when there are minor disturbances in network performance, especially for vehicular network in which terminals gather information and perform handoffs simultaneously. Decline in the targeted network performance after an influx of simultaneous handoffs only causes a new round of massive handoffs. A comparison between conventional and ideal handoff strategies is shown in Fig. 1. The inappropriate strategy makes convergence impossible.

The present study established a distributed vertical handoff strategy for heterogeneous vehicular networks. We used an optimal game approach to formulate the network selection process as a competition model for the terminals in the related area competing for the network with better performance. We selected an equilibrium specially designed for the heterogeneous vehicular network as the solution of the game; this allows us to fully exploit the advantage of DSRC and avoid potential disturbance on LTE and other networks. Local network performance and remote terminal status are combined as the optimal object so that the system can converge quickly with few jitters and avoid ping-pong effects. To verify the proposed system's performance in the real-world variable performance scenarios, we built and tested a prototype at CAVTest. The major contributions of this paper can be summarized as follows:

(1) We address the issue that conventional network selection strategies may lead to unexpected massive handoffs as the network performance changes. (2) An optimal game approach is proposed to solve the network selection problem when the performance of some networks varies. Terminals compete to access the network with better performance; the game is repeated by a probabilistic selection in period until no terminals can achieve better payoff with individual handoff. (3) A prototype of the heterogeneous vehicular network consists of DSRC, LTE, and Wi-Fi, and was established at CAVTest, on which the proposed network selection algorithm and traditional MCDM method were tested and compared in detail.

The rest of this article is organized as follows. In section II, related work is reviewed. The problem statement and system model are proposed in section III. The optimal game and the formulation are described in section IV. The simulation and experimental tests are presented in section V. Conclusions are provided in section VI.

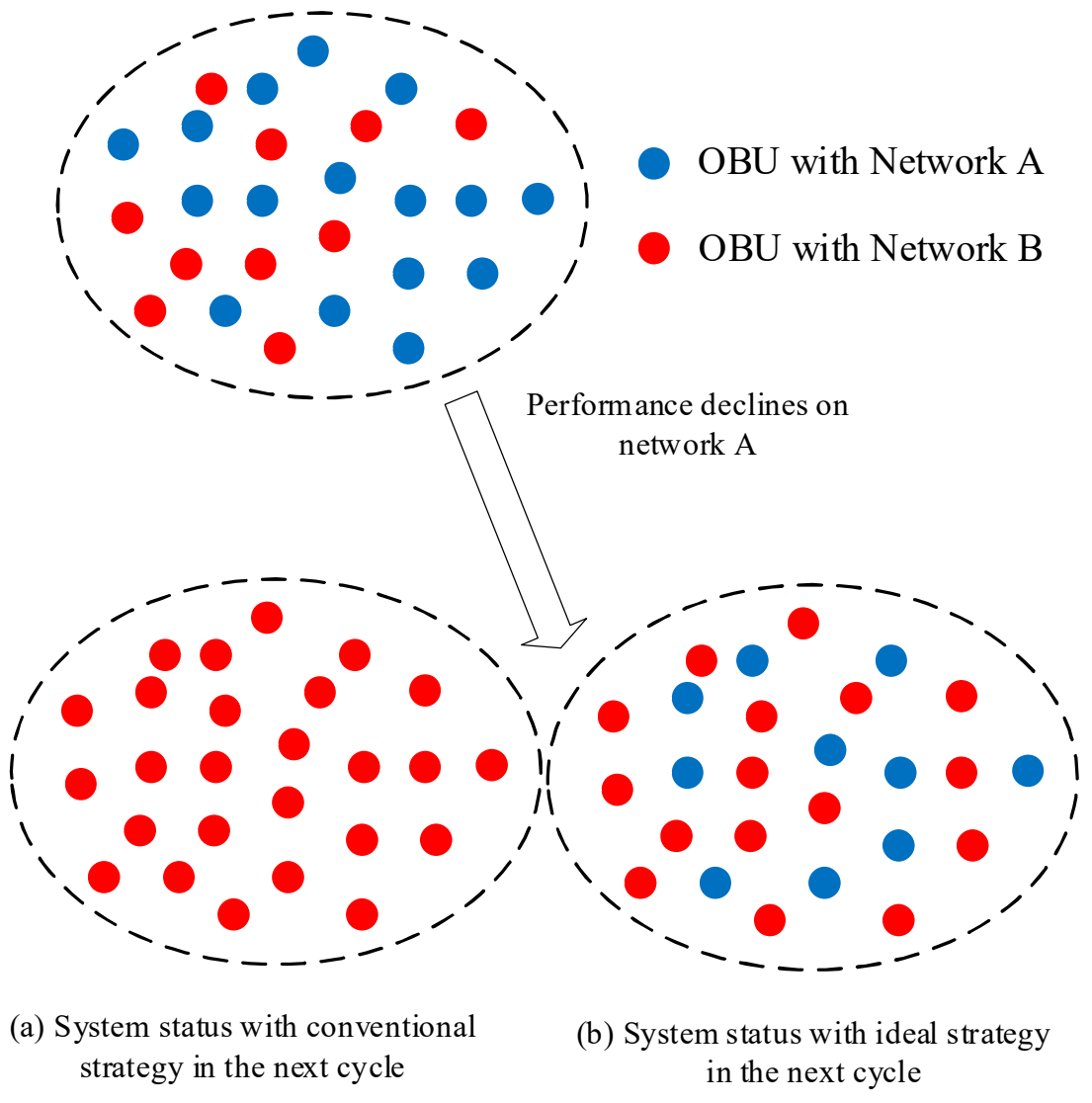

Fig. 1. System status between conventional strategy and ideal strategy when there is a performance decline in the network. The conventional strategy tends to switch all terminals from the network with declining performance, whereas the ideal strategy switches a portion of them to quickly restore stability.

## II. RELATED WORK

According to the best of our knowledge, most of conventional handoff strategies for heterogeneous vehicular network regard network performance as constant parameters.

Shafiee et al. [11] investigated optimal Vertical Hand Off (VHO) strategies in random inter-distance scenarios. Sepulcre et al. [12] presented a context-aware heterogeneous V2I communication technique that exploits context information to make intelligent decision regarding adequate communications technologies. Both of above methods only focus on the network bandwidth and are simulated in a low traffic volume scenario, so the handoff behavior will not affect the network performance. To improve both individual and system performance, Marquez-Barja et al. [13] proposed a vehicular heterogeneous vehicular network handoff algorithm empowered by the IEEE 802.21 standard, in which performance parameters of Wi-Fi, WiMAX, and UMTS were all fixed numbers. Tian et al. [14], inspired by the cellular gene network, proposed a dynamic and self-adaptive network selection method that enables terminals to dynamically select an appropriate access network. Wang et al. [15] presented a VHO method based on a self-selection decision tree of Internet of Vehicles (IoVs) to resolve issues with unstable network status and different user preferences. Fixed performance parameters were used in the simulation in the studies of Tian et al. [14] and Wang et al. [15]. Dey et al. [16] developed an application layer handoff method to enable traffic data collection and forward collision warning, and conducted a field study to validate their method. The handoff behavior was triggered by collision detection. The system had been verified in the field test, but the performance was relatively poor.

For these heterogeneous non-vehicular network systems, most researchers also regard network performance as constant parameters [17]–[19]. Niyato and Hossain [20] formulated competition among different areas with different networks as a dynamic evolutionary game, the solution to which they defined as evolutionary equilibrium. Multi-play game was used in the study of Niyato and Hossain [20], but the system was still simulated with fixed performance parameters. Liu et al. [21] found that network performance may degrade when certain terminals simultaneously handoff to the same network. However, the inherent of system performance reduction was overlooked. Liu et al. [22] used network parameters that randomly varied every time the network selection algorithm was executed to simulate its method, but this caused the network performance to be unrelated to the network selection behavior and did not comply with the real-world scenarios.

Based on the literature review, we found that both heterogeneous vehicular network and regular heterogeneous network selection methods lack of consideration on varying network performance, especially for the changing terminal number caused by network selection. The single-play network selection may cause stability issues in real world scenarios. The evolution-based game selection in [18] and [20] may help to solve the problem because of players' bounded rationality but features of vehicular network are not fully used. In vehicular network, terminals can obtain terminal number in the neighbor area, which will not happen in regular network. Apart from this, DSRC is dedicated to vehicular application, that makes the performance of DSRC is relatively predictable with the known terminal number. Both of these two features can be utilized in the vehicular heterogenous network selection. So, an optimal game approach for heterogeneous vehicular network selection with varying network performance is proposed to solve this problem. Both bounded rationality and vehicular network features are used in the optimal game to improve the performance of the presented method.

## III. Problem Statement and System Model

### A. Problem Statement

As discussed above, we built a heterogeneous vehicular network scenario with varying network performance to address the general problem of conventional handoff strategies. Suppose that there are several networks in a heterogeneous vehicular network system forming a set called $NetSet$. All OBUs form a set $MT$. An arbitrary OBU $i$ communicates with other OBUs only via network $j_i$, where $j_i \in NetSet$. The number of OBUs attached in network $j$ refers to $n_j$. The handoff decision is made every $T$ seconds. The targeted network of the heterogeneous vehicular network can be selected as follows:

$$j_i^* = \arg\max\{NetEva_{i,j} | j \in NetSet\} \ (\forall \mathrm{i} \in MT) \quad (1)$$

where $NetEva_{i,j}$ is the comprehensive evaluation of network $j$ determined by OBU $i$. Detailed algorithms to calculate $NetEva_{i,j}$ can be found in references [13], [14]. Most MCDMs conform to the following function:

$$NetEva_{i,j} = w_1 \times Performance_{i,j} + w_2 \times Context_{i,j} + w_3 \times Preference_{i,j} \quad (2)$$

where $Performance_{i,j}$, $Context_{i,j}$, and $Preference_{i,j}$ are the performance evaluation, context parameter evaluation, and user preference evaluation of network $j$ calculated by OBU $i$, respectively. As $Context_{i,j}$ and $Preference_{i,j}$ change relatively slowly, $Performance_{i,j}$ is the only significant evaluation. $Performance_{i,j}$ in network selection strategies is calculated by network performance parameters, such as delay, and PLR, which form a vector $\mathbb{x}_{i,j}$. These performance parameters and performance evaluations are both determined by the characteristics and context of the network, including the terminal number, barrier, signal noise, and so on, which form a state vector $\mathbb{s}_{i,j}$. $NetEva_{i,j}$ can be expressed as:

$$NetEva_{i,j} = Performance_{i,j} = G(\mathbb{x}_{i,j}) = F(\mathbb{s}_{i,j}) \quad (3)$$

where $G$ is treated as a function related to all the performance parameters. $F$ is treated as a function-related characteristic and context of the network. In the process of vertical handoff in the heterogeneous network, the terminal number of each network is directly decided by handoff strategy, making it distinct from the other contexts and features of the network. The comprehensive evaluation is also defined by:

$$NetEva_{i,j} = f_j(n_j) + F'(\mathbb{s}'_{i,j}) \quad (4)$$

where $f_j(n_j)$ is the relation function between the number of OBUs with network $j$ and network evaluation. The evaluation of the network degrades as the number of OBUs in the network increases [5]; this makes $f_j(n_j)$ a decreasing function.

We assume that there is a heterogeneous vehicular network system in which networks A and B are stable at time $t_0$. That means both networks share the same evaluation, and no terminals would switch to the other network, that is, $\forall m,n \in MT$, $NetEva_{m,A} = NetEva_{n,B}$. There are $n$ OBUs in the system, $g$ OBUs in network A and $h$ OBUs in network B. Suppose that there is a performance disturbance on network A at time $t_1$, making the comprehensive evaluation of network A worse than B by $\Delta E$. That is, at time of $t_1$, $\forall m,n \in MT$, the system follows:

$$\begin{cases} NetEva'_{n,A} = NetEva_{n,A} - \Delta E \\ NetEva'_{m,B} = NetEva_{m,B} \end{cases} \tag{5}$$

According to Eq. (1), all the OBUs attached in network A switch to network B. According to Eq. (4), $\forall m,n \in MT$,

$$\begin{cases} NetEva''_{n,A} = NetEva_{n,A} - \Delta E + f_A(0) - f_A(g) \\ NetEva''_{m,B} = NetEva_{m,B} + f_B(n) - f_B(h) \end{cases} \tag{6}$$

at time $t_1 + T$. This means that the performance of network A and B changes after the handoff behavior. The performance of network A increases by $f_A(0) - f_A(g)$ while the performance of network B decreased by $f_B(h) - f_B(n)$. If $NetEva''_{m,B} \geq NetEva''_{n,A}$, the system will remain stable after handoffs due to Eq. (1). If $NetEva''_{m,B} < NetEva''_{n,A}$, that is, $\Delta E < f_A(0) - f_A(g) + f_B(h) - f_B(n)$, all the OBUs will switch back to network A and the system will not return to stability. Because $f_A$ and $f_B$ are both decreasing functions and $g > 0$, $n > h$, $f_A(0) - f_A(g) + f_B(h) - f_B(n) > 0$. It is very likely that $f_A(0) - f_A(g) + f_B(h) - f_B(n) > \Delta E$ as $\Delta E > 0$. Thus, for a conventional heterogeneous vehicular network selection strategy, small variations in the comprehensive evaluation of existing vertical handoff strategies in heterogeneous vehicular networks may cause massive handoff behavior; this results in a strong disturbance in network performance and terminal attachment status, which we specifically designed the proposed strategy to prevent.

The aim of our vertical handoff strategy is to build stability in the heterogeneous vehicular network when there is disturbance in the network comprehensive evaluation. In the above-mentioned scenario, some OBUs with network A switch to network B so $\forall m,n \in MT$, $NetEva''_{m,B} = NetEva''_{n,A}$. If the system is stable after handoff decision, the number of OBUs that switch from A to B conform to the following:

$$f_A(g-s) - f_A(g) + f_B(h) - f_B(h+s) = \Delta E \tag{7}$$

where $\Delta E$ can be calculated easily because it is defined by the vertical handoff strategy. $g$ and $h$ are sensed by the OBU via wireless communication. Eq. (6) tells that to build a new stable system, the sum of the increased performance of network A and decreased performance of network B due to handoff should be equal to the disturbance on network A. Unfortunately, in the real-world scenario, it is impossible to ascertain $f_A$ and $f_B$ because there are excessive and random factors related to them (e.g., wireless channels, noise signals, barriers). We instead formulate the network selection of the heterogeneous vehicular network selection as a multi-play non-cooperative game. The solution is a specially designed equilibrium that includes consideration of network performance. The network evaluations are treated as payoffs related to the OBU choosing different networks.

The goals of the proposed optimal game approach for heterogeneous vehicular network selection are three-fold: (1) to take inconsistent network performance into consideration, especially the performance reduction caused by network handoffs; (2) to formulate the network selection process as an optimal game; and (3) to build up the game strategy as the characteristics of vehicular networks and corresponding environment are changing.

### B. *System Architecture*

The proposed heterogeneous vehicular network system consists of DSRC, LTE, and Wi-Fi, among which DSRC is the default network, as shown in Fig. 2. Network coverage and optimal states are not specifically required but it must ensure that at least one network is available. The network terminal like the OBU can connect to all three networks and perform the software-based handoffs individually. OBUs broadcast their own Basic Safety Message (BSM) every 0.1 s with the chosen network. Each message consists of the time stamp, identification, GPS positioning data, speed of movement, direction, and other relevant information of the source OBU. The broadcasting rate exploits the Coordinated Universal Time (UTC) provided by GPS as a global time reference. There is no inherent mechanism for broadcasting information to neighbor terminals designed in LTE and Wi-Fi, so the Integrated Inter-System Architecture (IISA) [23] is introduced to the system. The Interworking Decision Engine (IDE) is treated as a centralized control entity in IISA. The IDE receives all the BSM broadcast requests on LTE and Wi-Fi and retransmits the BSMs to the other OBUs. This also makes LTE and Wi-Fi terminals always on-line so that OBUs can receive information for LTE and Wi-Fi even if they do not use them to transmit information.

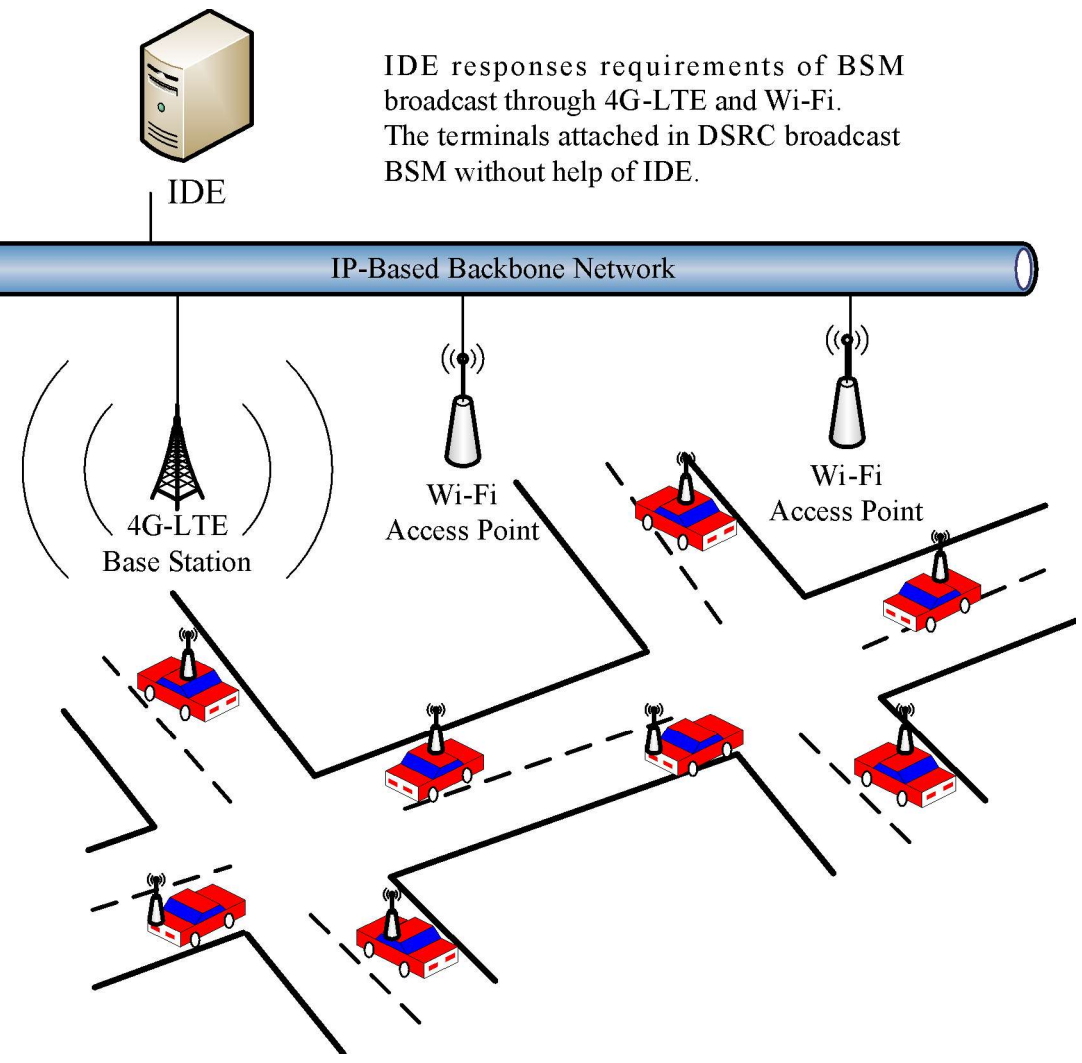

Fig. 2. Heterogeneous network system framework. OBUs can access all three networks: DSRC, LTE, and Wi-Fi.

In the network selection process, terminals collect network performance indicators, including propagation delay, PLR, and jitters. The network in our system has varying performance, so the available bandwidth cannot be measured. We calculate indicators from the information in the period-broadcasted BSMs. In every BSM cycle, the OBU first processes the information gathered in the last cycle, calculates the network evaluation, then makes the selection and broadcasts the BSM in this cycle with the chosen network. The process is repeated in the next cycle. We assume that all the terminals in the model share the same physical and network environment, so the information sent by every OBU can and should be obtained by any other OBU.

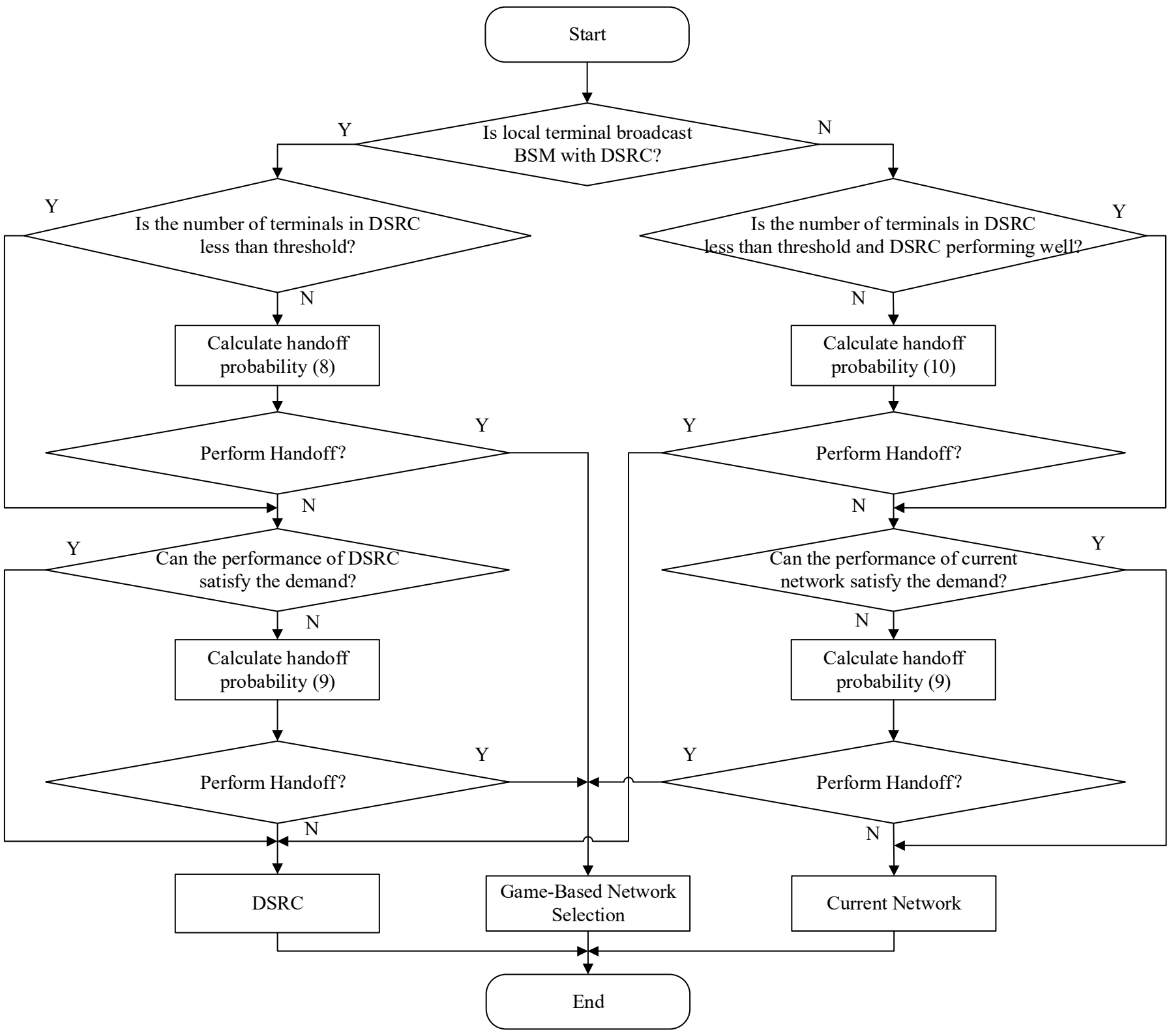

Fig. 3. Probabilistic network selection. An excess of terminals and/or performance degradation can trigger the game-based network selection.

## IV. Optimal Game and Formulation

### *A. A Brief Introduction to Game Theory*

Game theory is often applied to studies on the strategic interaction between agents in modeling optimization problems [24]. A set of players, their sets of strategies, corresponding payoffs, and the solution of all plays together form the "game". There are single-play, multi-play, cooperative, and non-cooperative games. In the single-play game, players choose the strategy that immediately leads to the desired solution; in the multi-play game, players can achieve the solution step-by-step through the information they obtain after each play. In the cooperative game, players select their strategy with consideration of other players. In the non-cooperative game, players do not take any joint consideration of other players. If no players can improve their payoffs without deviating from other players' strategies, the solution is a "Nash equilibrium"; this is the most popular solution.

Network performance declines as the number of terminals increases [5]. In the heterogeneous vehicular network, every handoff behavior can impact the performance of the target network. Any games designed to reach equilibrium immediately come with stability issues, as discussed in Section II. The traditional game theory must be modified to adapt to heterogeneous vehicular networks with varying performance.

### *B. Optimal Game Formulation for Heterogeneous Vehicular Network Selection*

The characteristics of the proposed heterogeneous vehicular network selection system with varying performance include (1) variable network performance according to the number of terminals, (2) knowledge of all the performance parameters of the available networks, and (3) knowledge of the number of surrounding terminals. We exploited these characteristics to build an optimal game to formulate the heterogeneous vehicular network selection process.

- *Players*: In our game, the player is the OBU that can access the heterogeneous network and has the ability to decide the network to which it should connect. The OBUs choose networks with the best payoffs (i.e., network evaluation).
- *Strategy*: The strategy of players encompasses all the available networks in the environment. Again, in our system, they are respectively DSRC, LTE, and Wi-Fi.
- *Payoff*: The network evaluation is the payoff. We use a comprehensive network evaluation including propagation delay, PLR, and jitter as the parameters to calculate the evaluation.

Player characteristics were already discussed in section III. Below, we introduce the strategy and payoff in detail.

We use DSRC, LTE, and Wi-Fi to set up a heterogeneous vehicular network system. To minimize switching behavior, DSRC is set as the default network. As opposed to the others, DSRC transmits data through broadcasting and adapts to the dynamic environment in the vehicular network. Its propagation delay is relatively low, which is suitable for safety applications of connected vehicles. Non-transportation applications on LTE and Wi-Fi, such as File Transfer Protocol (FTP) and streaming media, are likely to cause network instability, which impacts the overall performance. We built our algorithm to regard DSRC as the initial network to remedy this. A handoff is triggered when DSRC performance is unreliable, or the number of DSRC terminals is excessive according to its Media Access Control (MAC) layer protocol. To perform a multi-play game, the ideal of boundary rationality in the study of Niyato and Hossain [20] is used in our approach. Based on that, we proposed that a probabilistic handoff strategy can drive terminals to converge gradually to prevent instability.

When there are too many OBUs transmitting information with DSRC, the OBUs falling outside of the ideal amount need to switch to other networks. For OBU $i$ that currently uses DSRC to transform information, suppose that BSMs of $x$ OBUs are received through DSRC within the most recent three BSM cycles. This avoids underestimating the number of DSRC-broadcasting OBUs because of packet loss. Parameter $n_{exp}$ denotes the expected maximum DSRC terminals. If $x \leq n_{exp}$, the terminal does not do anything. If $x > n_{exp}$, the current terminal switches to a non-DSRC network at the probability of:

$$P_{switc} = \rho \frac{x - n_{exp} + 1}{x + 1} \quad (8)$$

where $\rho$ is the probability correction parameter. The probabilistic handoff strategy is to switch the redundant DSRC terminals to other networks. When $\rho = 1$, the system attempts to adjust the number of DSRC terminals to the ideal number immediately. This probably leads to ping-pong effects because all the terminals generate random numbers individually. We set $\rho < 1$ to prevent such effects.

Some performance parameters of connected vehicles are introduced to resolve issues with DSRC performance degradation. $F_{delay}$, $F_{PLR}$, and $F_{jit}$ represent the maximum acceptable propagation delay, maximum PLR, and jitter, respectively. If two or more performance parameters exceed the maximum acceptable parameters in a given BSM cycle, the network does not meet the performance requirements in that cycle, and OBUs try to switch to another network with better performance. Suppose that OBU $i$ receives BSMs from $x$ OBUs through the current network in the latest three BSM cycles. We want to make sure that the longer the network performance is unacceptable, the larger the probability will be that a terminal will switch to another network, so the performance reduction counter $c$ is introduced to the OBU. For each cycle in which the current network cannot meet the performance requirements, $c = c + 1$ and the OBU switches to another network at the following probability:

$$P_{switch} = \min(\frac{\sigma c}{x}, 1) \quad (9)$$

where $\sigma$ is the probability correction parameter. The proposed handoff possibility is to switch part of current network terminals to the other networks. The number of the switched terminals in this strategy is unrelated to the number of the attached terminals in the current network because it has been considered in Eq. (8). With Eq. (9), the OBU is more likely to connect another network when the current network cannot meet the performance requirements for a longer time. In every BSM cycle, the mean number of handoff terminal is $\sigma c$. When the current network meets the performance requirements in a BSM cycle, $c = c/2$, the OBU will not switch to another network.

The OBU $i$ broadcasting BSMs with a non-DSRC network receives BSMs of $x$ terminals through DSRC and BSMs of $x'$ terminals through the non-DSRC network in the latest three BSM cycles. $n_{exp}$ denotes the maximum acceptable DSRC terminal quantity as in (8). The OBU switches to DSRC at the probability of:

$$P_{switc} = \min(\rho \frac{n_{exp} - x}{x' + 1}, \rho) \quad (10)$$

where $\rho$ is the probability correction parameter, and we set $\rho < 1$ to minimize the ping-pong effects. This causes the non-DSRC terminals to switch back to DSRC to maximize the use of DSRC. A flowchart of the whole strategy is shown in Fig. 3. For the terminal attached in DSRC, it first checks if there are too many terminals attached in DSRC. If so, the terminal calculates the handoff probability according to Eq. (8) and decides whether to switch with the game-based approach. If a terminal does not perform a handoff in the first part, it checks if the performance of DSRC satisfies the application demand. If not, the terminal calculates the handoff probability according to Eq. (9) and decides whether it will execute the handoff. For a terminal attached in a non-DSRC network, it first checks whether the DSRC network can load more terminals, that is, whether the performance of the DSRC network satisfies the applications and the number of terminals attached in is below the threshold. If so, the terminal calculates the handoff probability and decides whether to switch using the game-based network selection. Otherwise, the terminal would check whether the current network can satisfy the application demands. If not, the terminal will calculate the handoff probability according to Eq. (9). After all the above, if there is no handoff decision to be made, the terminal will stay in the current network to transmit messages.

The payoff is the evaluation of each network. We use MCDM, the most commonly used method to evaluate networks in our system. In the proposed optimal game, the network evaluation is regarded as the payoff of network selection. Because we are not intending to improve the methodology of MCDM, only network performance is taken into consideration in network evaluation. In the present system, available bandwidth cannot be measured directly because the proposed strategy works with varying network performance. Propagation delay, PLR, and jitter are the performance parameters in this system. All the parameters can be obtained by BSMs. For network $j$, $f_{jdelay}$ denotes propagation delay, $f_{jPLR}$ denotes PLR, and $f_{jjit}$ denotes delay jitter. They can be obtained by:

$$f_{jdelay} = \frac{\sum_{i=1}^{n_j}(t_i - t_0)}{n_j} \quad (11)$$

$$f_{jPLR} = \frac{n_{jamount} - n_j}{nj} \quad (12)$$

$$f_{jjit} = \frac{\sum_{i=1}^{n_i}|\tau_{i,t} - \tau_{i,t-1}|}{n_j} \quad (13)$$

where $t_0$ denotes the time that the BSM is produced, $t_i$ is the time that the local terminal receives BSM from

vehicle$i$, $n_{jamount}$ is the number of vehicles that the local OBU perceives through network $j$ in the last second, $n_j$ is the number of vehicles perceived through BSMs within the current cycle by network $j$, $\tau_{i,t}$ is the propagation delay between the local terminal and remote terminal $i$ of the current BSM cycle, and $\tau_{i,t-1}$ is the propagation delay between the local terminal and remote terminal $i$ of the last BSM cycle. The utility variables can be calculated with a simple linear normalization method to nondimensionalize the different performance attributes:

$$f'_{jdelay} = \frac{F_{delay} - f_{jdelay}}{F_{delay}} \tag{14}$$

$$f'_{jPLR} = \frac{F_{PLR} - f_{jPLR}}{F_{PLR}} \tag{15}$$

$$f'_{jjit} = \frac{F_{jit} - f_{jjit}}{F_{jit}} \tag{16}$$

where $F_{delay}$, $F_{PLR}$, $F_{jit}$ are reference parameters of propagation delay, PLR and jitter respectively. The network evaluation of OBU $i$ and network $j$ can be obtained as follows:

$$NetEva_{i,j} = w_1 f'_{jdelay} + w_2 f'_{jPLR} + w_3 f'_{jjit} \tag{17}$$

where $w_1$, $w_2$,and $w_3$ are the weight coefficients of propagation delay, PLR, and jitter, respectively.

## V. Simulation and Field Test

### A. Simulation

We built a MATLAB Simulink model of the proposed game approach. Several scenarios were simulated, including the step response when DSRC terminals are too few and when there is influence of traffic disturbances. Parameters used in the simulation are shown in TABLE II. The ideal number of DSRC and initial terminal numbers are decided according to the network performance of each network to ensure that no single network, but a coordinated heterogeneous network system, can load these terminals. Decreasing convex functions are used to obtain the network performance based on terminal number. Observation results include the terminal number of all three networks, the network performance, and the handoff terminal number in each cycle which reflects ping-pong effect.

Fig. 4(a) and 4(b) show the results of the step response on both optimal game approach and conventional MCDM. Fig. 4(c) illustrates the network performance of DSRC, LTE, and Wi-Fi for the proposed optimal game approach, whereas Fig. 4(d) shows the average network performance of all terminals for both optimal game approach and conventional MCDM. It can be concluded that the proposed method drives terminals to a stable state, i.e., equilibrium, within 1 second. All three networks are fully utilized. Conventional MCDM cannot achieve stability during the whole simulation process. All terminals trying to switch to the network with the best evaluation causes massive handoff and ping-pong effect. From Fig. 4(b), we also observe that there are no terminals attached in LTE because according to our performance function, compared with other networks, LTE has the worst performance when there is only one terminal attached in it. So, terminals with conventional MCDM can always find a better network other than LTE, where no terminals are attached all the time. In Fig.

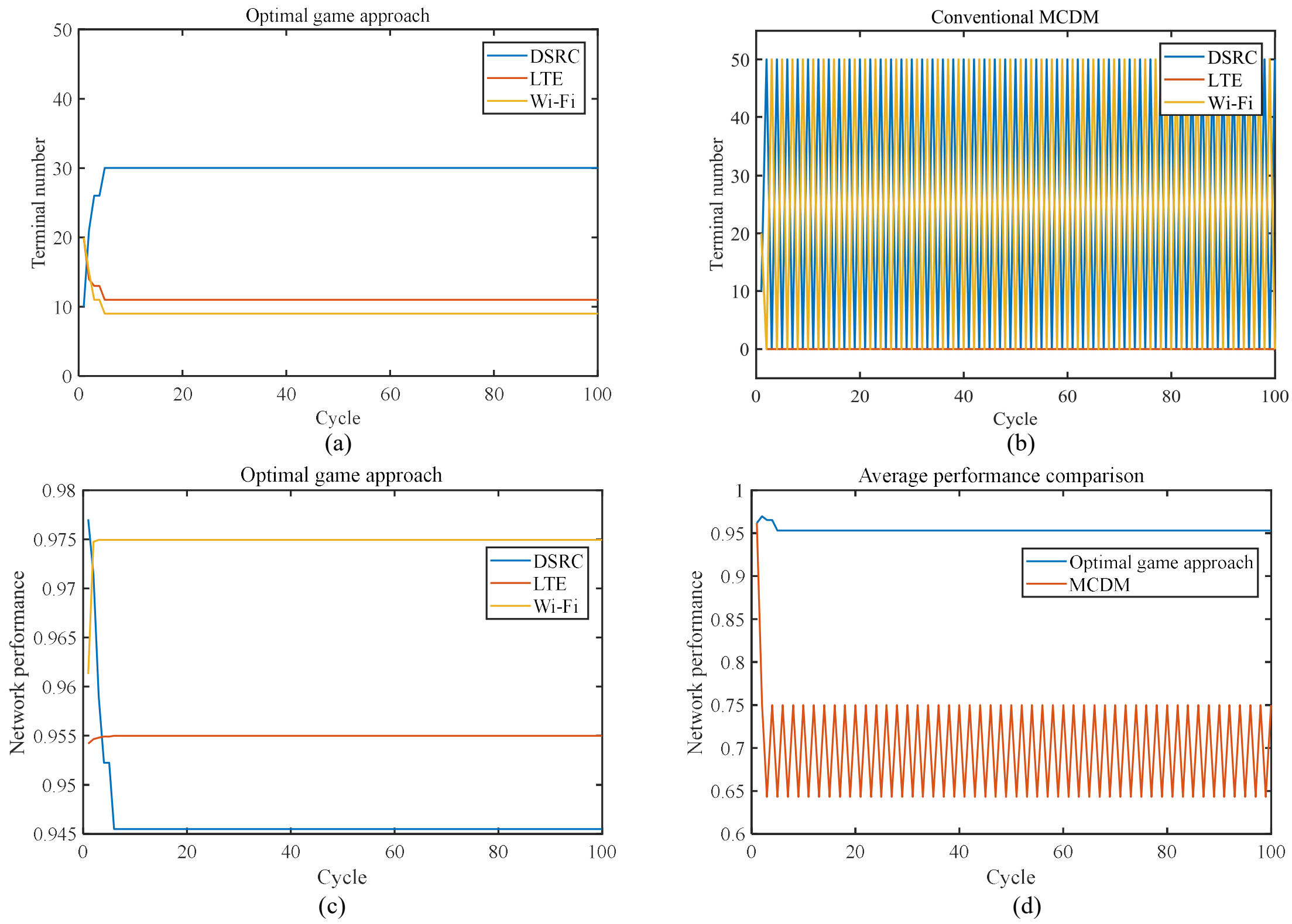

Fig. 4. Simulation results with step response. (a), (b) Simulation results of step response on both optimal game approach and conventional MCDM. (c) Network performance of DSRC, LTE, and Wi-Fi for the proposed optimal game approach. (d) Average network performance of all terminals for both optimal game approach and conventional MCDM.

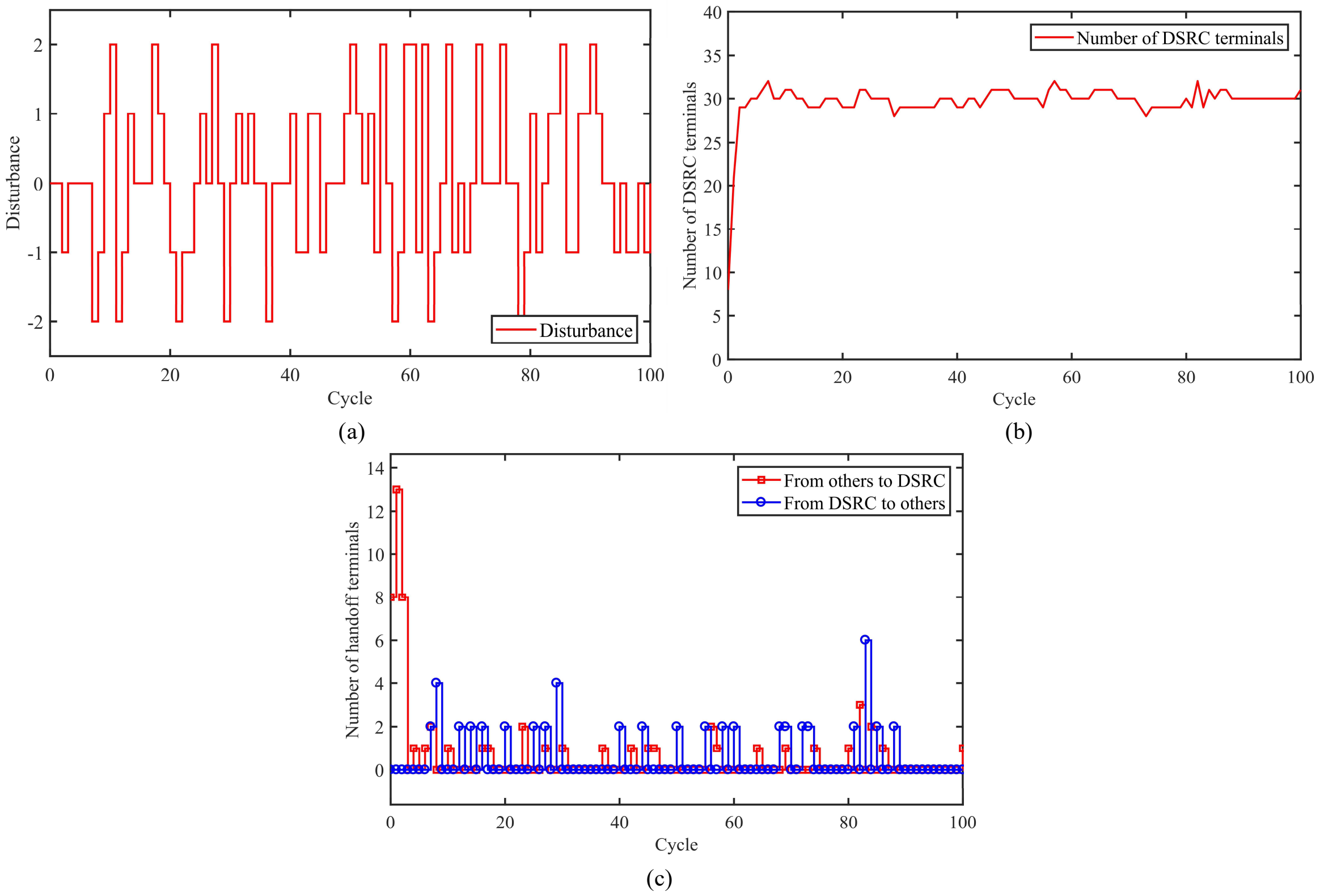

Fig. 5. Simulation results with disturbance. (a) Input noise. (b) Changes of the DSRC terminal number. (c) Number of the terminals which execute handoffs between DSRC and other networks.

4(c), it can be observed that DSRC is overflowed because both LTE and Wi-Fi have better performance evaluation than it after equilibrium. This is a trade-off produced by our proposed method. To maximize the use of DSRC and avoid potential instability of LTE and Wi-Fi, it is possible for a terminal to choose DSRC even if its performance is not as good as the others. In Fig. 4(d), the proposed optimal game approach dominates the MCDM during the whole simulation period on the average performance of all terminals. The small performance peak of proposed method illustrates that during the converging process, the overall performance is better than the performance after convergence. It is also the trade that we made to maximize the use of DSRC and guarantee the stability of the vehicular network.

TABLE II
PARAMETERS FOR SIMULATION

| Items | Step Response Simulation | Disturbance Simulation |
|---|---|---|
| Ideal number of DSRC-broadcasting OBUs ($n_{exp}$) | 30 | |
| Total number of OBUs | 50 | |
| Initial number of DSRC, LTE and Wi-Fi terminals | 10, 20, 20 | 0, 25, 25 |
| Probability correction parameter ($\rho, \sigma$) | 0.5, 0.5 | |
| Noise amplitude | / | 2 |
| Noise frequency | / | 10Hz |

Fig. 5 shows the network convergence and switching times when there is a noise on DSRC-broadcasting terminal number. Fig. 5(a) refers to the disturbance on the number of vehicles within the wireless communication coverage area. The frequency of the noise was set to 10 Hz, and the amplitude to 2 vehicles in this simulation and the random noise of each vehicle was independent from others. The number of DSRC terminals is relatively stable with a maximum offset of 2 vehicles as shown in Fig. 5(b), after the first convergence of the system. Disturbance is observed in Fig. 5(c) because there were several handoffs in every BSM cycle in the system. As stable number of DSRC terminals in Fig. 5(b) shows, the system still appears to be reliable because an individual handoff did not affect the overall network performance. Taken together, our simulation results demonstrate that the system retains stability and restrains the ping-pong effects effectively with high-frequency scrambling.

We found that the proposed optimal game approach is effective in coordinating the appropriate switching of terminals in the heterogeneous vehicular network. The algorithm can quickly respond to the problem of DSRC message broadcasting congestion and resource idling in the heterogeneous vehicular network, enabling it to achieve convergence and overcome ping-pong effects effectively. We also found that the algorithm can adapt to varying network performance and system disturbances.

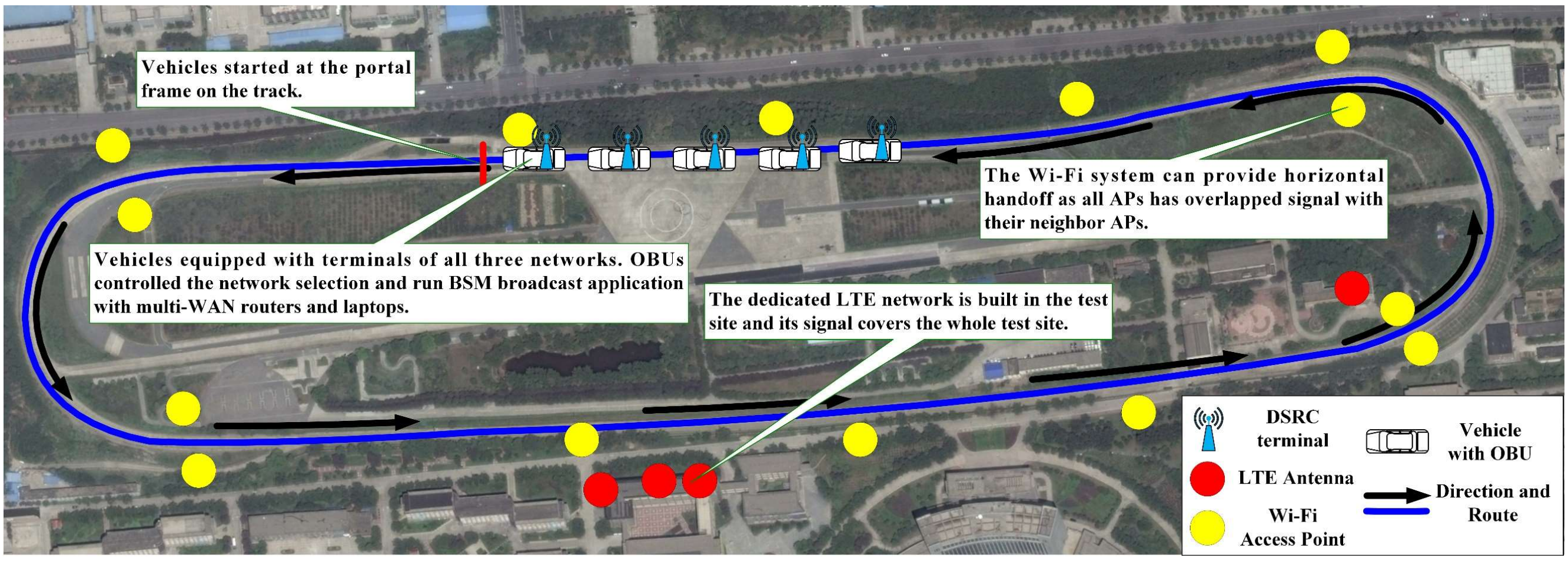

Fig. 6. The test scenario of the proposed prototype.

## B. *Field Test*

To verify the advantage of the proposed strategies, we test the system in a high-density scenario. Because of the limitations of test conditions, and limited network terminals, we could not test the algorithm in a high-density terminal scenario. Instead, we used five sets of DSRC, LTE, and Wi-Fi devices to imitate a heterogeneous network system with 50 OBUs. The broadcasting rate exploits UTC provided by GPS as a global time reference. The broadcasting frequency of single-equipment BSMs was increased to simulate a scenario with higher vehicle density. Ten BSMs were broadcasted in each BSM cycle, and five terminals were used to simulate 50 terminals. Each packet independently completed the optimal game approach with its own state variable. It is worth mentioning that network performance in such an approximate system is better than that in the corresponding real scenario. When there are 50 terminals, the channel contention will be more intense, and the network performance may degrade more. Therefore, our test only illustrated the effectiveness of the algorithm to a certain extent and merits further research. The parameters we used are shown in TABLE III. The test scenario is shown in Fig. 6.

We used five vehicles in the test with 25 m average spacing between 2 consecutive vehicles. All the vehicles moved counter-clockwise around the test bed for two rounds at a speed of 60 km/h, which took about 5 minutes. The data of all 50 virtual OBUs were recorded including network performance indicators, such as propagation delay, PLR, jitter, and the attached network of every virtual OBU. The proposed strategy and a strategy only using single-play MCDM were both tested for comparison. The test results are shown in Fig. 7.

Fig. 7(a) shows that the proposed vertical handoff strategy significantly improves the network on all three performance indicators, including delay, jitter, and PLR, in the high-density scenario. The PLR of the strategy without the proposed game approach was relatively high because the first vehicle and the last vehicle in the test were working with a shielded DSRC signal and could not transmit BSMs to each other directly through DSRC very well. The proposed strategy provided a reliable LTE connection for the virtual terminals on the first and last vehicles, so the PLR was much lower. With the proposed optimal game approach, the total handoff time is 10% less than the counterpart of conventional MCDM method. The handoff probability of each terminal is as low as 1.67%, which means that the terminals make a handoff decision every 6 seconds on average, whereas terminals working with MCDM without optimal game approach switch to another network every 0.5 second, on average.

Fig. 7(b) shows variations in the number of DSRC terminals. In our test scenario, DSRC had lower propagation delay and higher PLR, whereas LTE had higher propagation delay and lower PLR. They shared similar network performance evaluations for this reason. The proposed strategy proved relatively stable with DSRC terminals numbering between 34 and 27. Although the proposed strategy is relatively stable, it did not achieve stability during the whole test for the below reasons:

1) The vehicles had different contexts in real-world scenarios and the performance of the network is always changing.
2) The number of DSRC-broadcasting terminals is obtained by BSMs, and the number could be underestimated due to packet loss.

TABLE III
PARAMETERS IN THE FIELD TEST

| Handoff Strategy | | DSRC | | LTE | | Wi-Fi | |
|---|---|---|---|---|---|---|---|
| $\rho$ | 0.5 | Tx. Power | 22 dBm | Duplex mode | TDD | Wireless standard | IEEE 802.11n |
| $\sigma$ | 0.5 | Freq. | 5.9 GHz | Freq. | 1.9 GHz | Freq. | 2.4 GHz |
| $n_{exp}$ | 30 | Channel BW | 10 MHz | Resource allocation | 2DL+Dw-PTS+2UL | Tx. power (AP) | 20 dBm |
| $F_{delay}, F_{PLR}, F_{jit}$ | 0.1 s, 5 %, 0.1 s | CCH: SCH | 1:1 | Tx. power (UE) | 10 dBm | Tx. power (UE) | 12 dBm |
| $w_1, w_2, w_3$ | 0.7, 0.2, 0.1 | Packet size (for all) | 300 Bytes | Tx. power (enodeB) | 30 dBm | MIMO | None |

3) To overcome the underestimation on the number of DSRC-broadcasting terminals, the number of vehicles of which the BSMs are received in the latest three cycles through DSRC is used in the proposed approach to estimate the actual terminal number. The number can be overestimated when there are terminals that switch to a non-DSRC network in the latest three cycles.

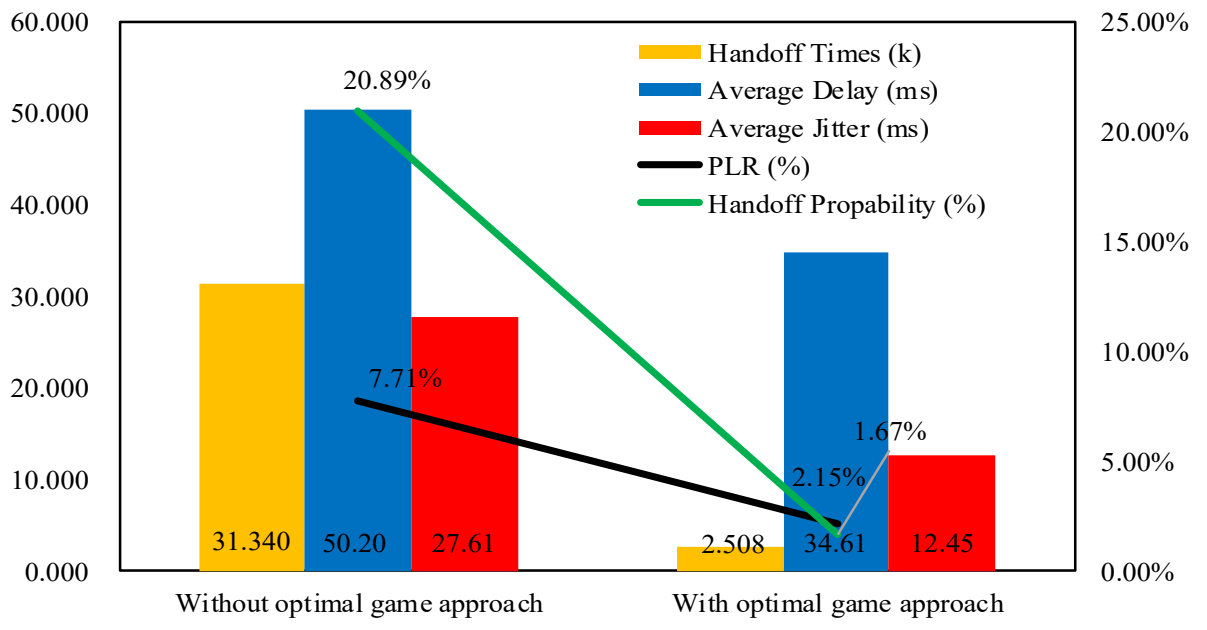

(a)

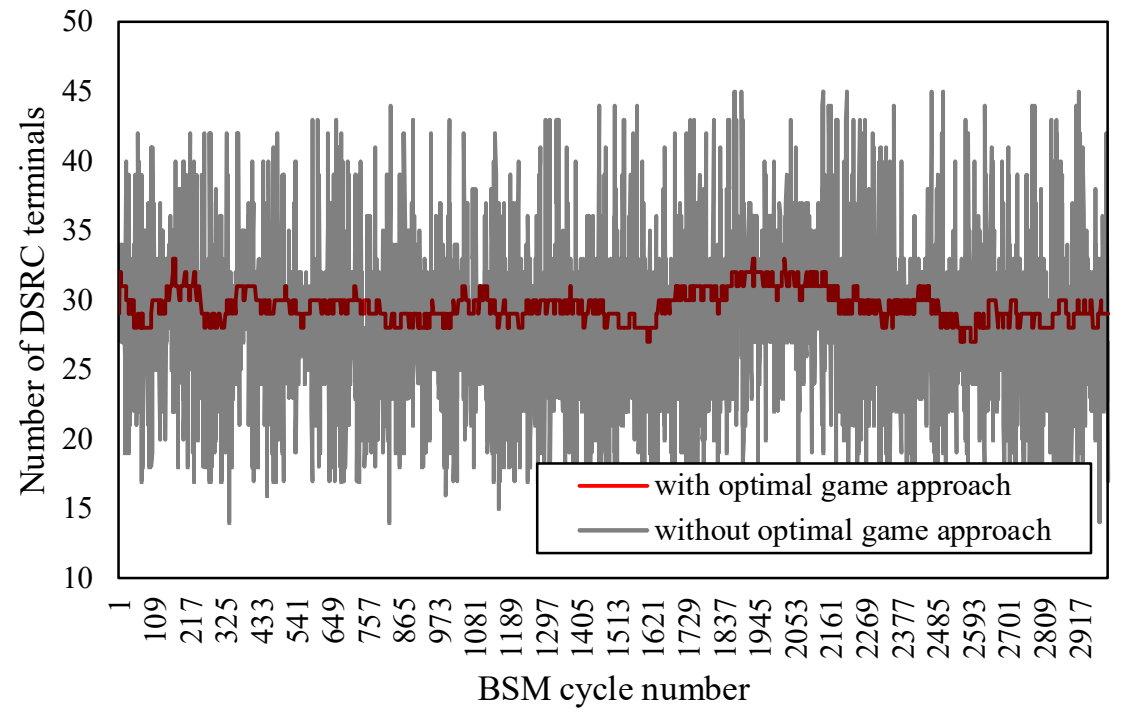

(b)

Fig. 7. (a) Network performance comparison between the proposed optimal game approach and conventional MCDM. (b) Number of DSRC terminals between the proposed approach and conventional MCDM.

## VI. Conclusions

The existing vertical handoff strategies for heterogeneous vehicular networks rarely consider the handoff effect on network performance. This paper proposed a novel optimal game approach for heterogeneous vehicular network selection. An optimal multi-play non-cooperative game model was established based on a careful review of the current vertical handoff problems and the actual characteristics of heterogeneous vehicular networks, which is more helpful when making decision on heterogeneous vehicular network selection according to varying network performances. MATLAB Simulink simulations were conducted to assess the strategy stability when with different input functions. We also built a prototype at CAVTest and compare the proposed method with a single-play algorithm. The testing results show that the proposed algorithm can overcome the identified problems on the single-play MCDM algorithm, making the proposed method more suitable for heterogeneous vehicular networks with varying performance.

Although the proposed strategy may outperform most of other existing handoff strategies in certain real-world scenarios, we were only able to test it in a relative ideal network environment with 50 simulated OBUs. Further open road tests will inevitably encounter some serious problems. For instance, the network performance is obtained by BSMs broadcasted by nearby OBUs which may lead to an underestimation of the total number of DSRC-broadcasting terminals. In addition, there have been many previous studies on vertical handoff selection strategies but few studies on system models or comprehensive evaluation methods specially for heterogeneous vehicular network. In the future, we will devote to study on the above two issues, which are more helpful for further optimizing the handoff strategies.

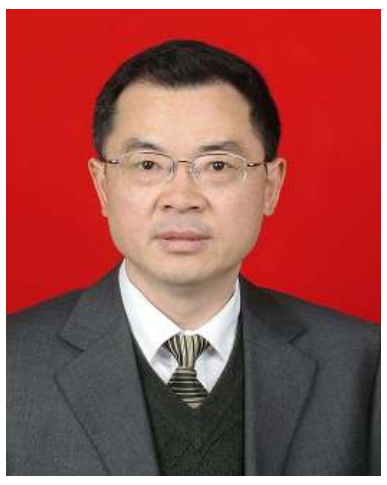

**Xiangmo Zhao** received the B.S degree from Chongqing University, China, in 1987, the M.S. and Ph.D. degree from Chang'an University, China, in 2002, 2005 respectively. He is currently a professor and the vice president of Chang'an University, China. He has authored or co-authored over 130 publications and received many technical awards for his contribution to the research and development on intelligent transportation systems. His research interests include Intelligent Transportation Systems, distributed computer networks, wireless communication and signal processing.

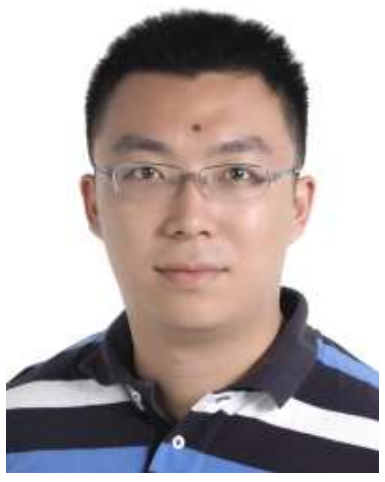

**Xiaochi Li** is a Ph.D. student in the School of Information Engineering, Chang'an University, China. He received his B.S. degree from Chongqing University, China, in 2012 and M.S. degree from Chang'an University, China, in 2015. His research focuses on the empirical study of connected vehicle and heterogeneous vehicular network.

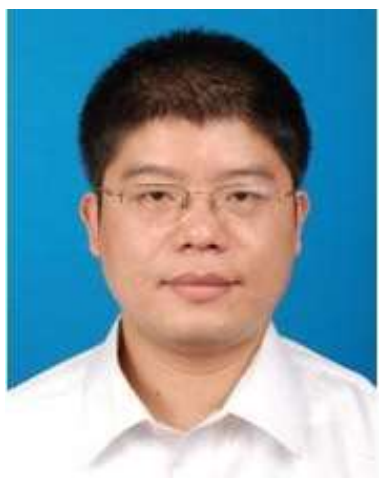

**Zhigang Xu** received his B.S. degree in automation, M.S. and Ph.D. degree in traffic information engineering and control from Chang'an University, China, in 2002, 2005, 2012, respectively. He is currently an associate professor in Chang'an University, China. His research focuses on intelligent diagnose on vehicle, connected and autonomous vehicle, and nondestructive examination on traffic infrastructure.

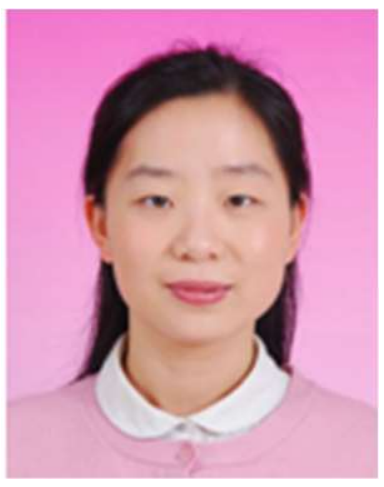

**Ting Chen** received her B.S. degree in communication engineering, and Ph.D. in information and communication engineering from Xidian University, China, in 2004 and 2011, respectively. She is currently an associate professor in Chang'an University, China. Her research focuses on wireless communication, Intelligent Transportation System, and pattern recognition.